%% file: sample-sigconf.tex
\documentclass[sigconf,screen]{acmart}

\usepackage[hide]{chato-notes}
\usepackage{comment} 
\usepackage{enumitem,kantlipsum}
\usepackage{multicol}

\def\BibTeX{{\rm B\kern-.05em{\sc i\kern-.025em b}\kern-.08emT\kern-.1667em\lower.7ex\hbox{E}\kern-.125emX}}

\setcopyright{none}
\acmConference[ComplexRec 2020]{Workshop on Recommendation in Complex Scenarios at the ACM RecSys Conference on Recommender Systems (RecSys 2020)}{22 September 2020}{Rio de Janeiro, Brazil}
\acmDOI{}
\acmISBN{}
\copyrightyear{2020. Copyright for the individual papers remains with the authors. Copying permitted for private and academic purposes. This volume is published and copyrighted by its editors}
\acmPrice{}
\begin{document}
\setlength{\abovedisplayskip}{3pt}
\setlength{\belowdisplayskip}{3pt}

%
\title[Scalable Recommendation of Wikipedia Articles to Editors]{Scalable Recommendation of Wikipedia Articles to Editors Using Representation Learning}

%

 \author{Oleksii Moskalenko}
 \affiliation{%
   \institution{Ukrainian Catholic University}
   \city{Lviv}
   \country{Ukraine}}
  
 \author{Denis Parra}
 \affiliation{
     \institution{Pontificia Universidad Catolica de Chile \& IMFD}
     \city{Santiago}
     \country{Chile}
 }
  \author{Diego Saez-Trumper}
 \affiliation{
     \institution{Wikimedia Foundation}
     \city{San Francisco}
     \country{USA}
 }


%

%
\begin{abstract}
\input{sections/abstract.tex}
\end{abstract}

%
%






\maketitle
\input{sections/010.tex}
\input{sections/020.tex}

\input{sections/030.tex}

\input{sections/040.tex}

\input{sections/050.tex}
\input{sections/060.tex}



\bibliographystyle{ACM-Reference-Format}
\bibliography{my}

\end{document}

%% file: sections/abstract.tex
Wikipedia is edited by volunteer editors around the world. Considering the large amount of existing content (\textit{e.g.} over 5M articles in English Wikipedia), deciding what to edit next can be difficult, both for experienced users that usually have a huge backlog of articles to prioritize, as well as for newcomers who that might need guidance in selecting the next article to contribute. Therefore, helping editors to find relevant articles should improve their performance and help in the retention of new editors. In this paper, we address the problem of recommending relevant articles to editors. To do this, we develop a scalable system on top of Graph Convolutional Networks and Doc2Vec, learning how to represent Wikipedia articles and deliver personalized recommendations for editors. We test our model on editors' histories, predicting their most recent edits based on their prior edits. We outperform competitive implicit-feedback collaborative-filtering methods such as WMRF based on ALS, as well as a traditional IR-method such as content-based filtering based on BM25. All of the data used on this paper is publicly available, including graph embeddings for Wikipedia articles, and we release our code to support replication of our experiments. Moreover, we contribute with a scalable implementation of a state-of-art graph embedding algorithm as current ones cannot efficiently handle the sheer size of the Wikipedia graph.

%% file: sections/010.tex
\section{Introduction}

Wikipedia is edited by hundreds of thousands of volunteers around the world. While the level of expertise, motivations, and time dedicated to that task varies among users, most of them experience challenges in deciding which articles to edit next. For example, many experienced users have huge backlogs\footnote{\url{https://en.wikipedia.org/wiki/Wikipedia:Backlog}} of work with a large number of articles to improve or review. Prioritizing the articles in this backlog, as via a personalized article ranking system, would potentially be of great help for these editors. On the other hand, newcomers might experience difficulties deciding what do to after their first contribution, as evidenced by the many efforts to understand how to improve the retention of newcomers~\cite{newcomers1,newcomers2}. 

While previous work on recommendations in Wikipedia has focused on finding articles for translation~\cite{wulczyn2016growing} or content to be added to existing articles~\cite{piccardi2018structuring}, there are still important unsolved problems: i) creating a \textit{scalable} recommender system that can deal efficiently with the large number of Wikipedia editors (over 400K monthly just in the English Wikipedia) and articles~\cite{wikistats2} ii) having good coverage of articles beyond just the most popular articles, and, iii) being able to provide good recommendations for newcomers, facing the classical user cold-start problem~\cite{schein2002methods}. 

To address these problems, we have created an efficient and scalable implementation of a state-of-art convolutional graph embedding algorithm~\cite{graphsage} that is able to deal with the large Wikipedia article graph. We combine this with a document embedding model that allows us to learn representations for articles and editors, and does not require retraining when new users are added in the system. With only a few edited articles then, the system is able to produce personalized recommendations, similar to Youtube deep recommendation model~\cite{Covington:2016:DNN:2959100.2959190}. We test our algorithm in English Wikipedia (the largest one with almost 6 million articles
showing that we can overcame well-established content-based filtering methods as well as collaborative filtering approaches. Moreover, we evaluate our recommendation measuring the top-100 items, to support a robust evaluation against popularity bias~\cite{valcarce2018robustness}. 

In summary, the main contributions of this paper are: \emph{(i)} Introduce a model which learns representations (graph and content-based) of Wikipedia articles and makes personalized recommendations to editors; \emph{(ii)} Evaluate it with a large corpus, comparing with competitive baselines 
\emph{(iii)} and release a scalable implementation of GraphSage, a state-of-art graph embedding system, that in previous implementations was unable to deal with the large graph of Wikipedia page\footnote{\url{https://github.com/digitalTranshumant/WikiRecNet-ComplexRec2020}}.

%% file: sections/020.tex
\begin{figure*}[t!]
\includegraphics[width=0.88\linewidth]{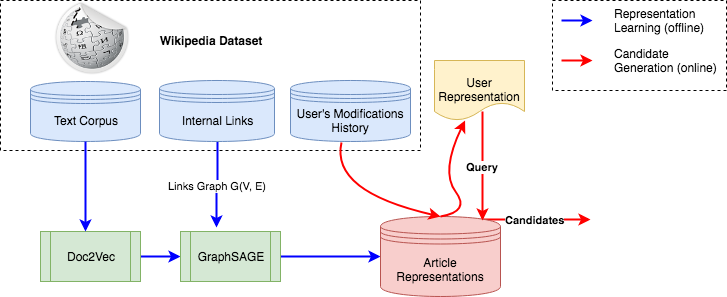}
\caption{Flow of Candidate Generation for Wikipedia articles recommendation: Doc2Vec embeddings are trained on Wiki Text Corpus and then passed as input into GraphSAGE model. Received articles' representations are then used in Nearest Neighbors Search to produce candidates}

\label{fig:candidate-generation}
\end{figure*}
\section{Related work}
\label{sec:ref}

There are several projects trying to solve the task of recommending items to users at real-world scales of millions of users and millions of items. For instance, Ying et al. for Pinterest ~\cite{Ying2018GraphCN} created an extension of GraphSAGE~\cite{graphsage}, a type of Graph Convolutional Network~\cite{gcn} (GCN); researchers at YouTube~\cite{Covington:2016:DNN:2959100.2959190} built a system based on regular deep neural networks that jointly learns users' and items' representations from users' previous history of views. However, in both examples, the model  learns in a supervised setup, whereas we lack a sufficiently comprehensive dataset of previous interactions because 94\% of Wikipedia contributors are associated with less that 10 interactions in last 3 years~\cite{wikistats2}. eBay's recommendation system covers a similar gap by using TF-IDF for similar item search, which does not require training~\cite{Brovman:2016:OSI:2959100.2959166}. 

On the document representation task, we can highlight several approaches: Doc2Vec\cite{doc2vec} is method for obtaining content-based representations of paragraph or longer text in vector space. However, one main advantage of our dataset is the availability of structural knowledge~\cite{consonni2019wikilinkgraphs} - i.e. links among articles that could potentially tell more about the article beyond its content. Those links can be represented as a graph, where nodes are articles and edges are links between them. Thus, the task of learning document representations can be transformed into learning the representation of a node in the graph. Node2vec~\cite{GroverL16} is a recent approach to learn such a representation. However, its scalability is still limited~\cite{fastnode2vec} and the main drawback for our use case is the necessity of full retraining after changes in the structure of the graph. Node2vec also omit the content part of articles (node features), which is a huge part of our dataset.

GCN~\cite{gcn, NIPS2015_5954} are a recent approach to solve many machine learning tasks like classification or clustering of a graph's nodes via a message-passing architecture that uses shared filters on each pass. It combines initial node features and structural knowledge to learn comprehensive representations of the nodes. However, the original GCN  architecture is still not applicable to large-scale graphs because it implies operations with a full adjacency matrix of the graph. To tackle these limitations, GraphSAGE model was introduced~\cite{graphsage} in a way that only some fixed-sized sample of neighbors is utilized on the convolutional Layer. Because of the fixed-size samples, we also have fixed-sized weights that are generalized and could be applied to a new, unknown part of the graph or even completely different graph. Thus, with inductive learning, we can train the model on a sub-graph, which means less computation resources are required, and evaluate generalization on the full graph.

%% file: sections/030.tex
\section{WikiRecNet Description}
\label{sec:wikirec}
Here we introduce \textit{WikiRecNet}, a scalable system for providing personalized article recommendations in Wikipedia,  built on top of GCN and Doc2Vec. The  design of our solution is inspired by a classic \textit{Information Retrieval} architecture. First we represent users by the articles that they edited, then  we  generate a list of candidates from the article pool by comparing that user representation with the article representations. Next, we sort the article candidates accordingly to the user preferences and generate a list of \textit{top-n best candidates} recommendation.

\subsection{Article and user representation} The primary challenge for our system is producing good user and article representations. It is an especially big problem for user representation since most of Wikipedia contributors do not fill any additional information about themselves except their login credentials \footnote{\url{https://en.wikipedia.org/wiki/Wikipedia:Wikipedia_is_anonymous}}, and around 28\% of all revisions in our  English Wikipedia dataset, are done by anonymous users~\cite{wikistats2}. The only useful information that could uniquely characterize the user is the history of his editions. Hence, most of our efforts were dedicated to learning articles' representations, and then representing the user based on the articles edited. One effective approaches to construct good user and item representations is to learn them with recommendation supervision
~\cite{Covington:2016:DNN:2959100.2959190,Ying2018GraphCN}. However, it is not possible to follow this approach due to the lack of the required comprehensive-enough dataset of previous interactions.  History of users editions in Wikipedia is far from exhaustive (88\% of users of English Wikipedia have done less than 5 major editions~\cite{wikistats2}) and too sparse, in a way that it is hard to model user's area of interest. Therefore, the additional challenge is to conduct representation learning~\cite{bengio2013representation} in an unsupervised way in relation to our final task.

\begin{figure}
	\includegraphics[width=0.9\columnwidth]{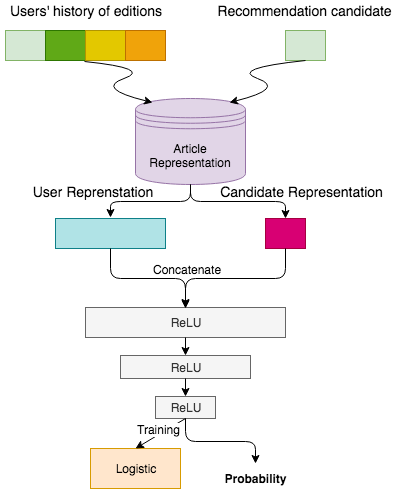}
	\caption{Candidate Ranking: user  history along with candidate are passed through articles' representation database (Embedding Layer) and then through several fully-connected layers to train in the log-regression setup.}
	\label{fig:ranking} 	
\end{figure}

\begin{table}
	\caption{Performance of different algorithms for K-NN search. All tests were conducted with English Wikipedia articles ($|V|=5,251,875$). Setup is measured in seconds. Secs. req. means seconds per request.}
	\centering      
	\setlength\tabcolsep{3pt}.      
	\begin{tabular}{l|c|c|c|c}\hline
		Algorithm & Setup & Secs./req. & Recall & MRR  \\
		\hline
		Exact search & 3.91 & 0.81 & 0.224 & 0.0220 \\
		IVF & 207.02 & 0.07 & 0.206 & 0.0212 \\
		HNSW & 232.68 & 0.04 & 0.224 & 0.0220 \\
		LSH & 472.31 & 0.15 & 0.215 & 0.0219\\
	\end{tabular}
	\label{tab:my_label}
\end{table}

\begin{table}
	\caption{Specifications of built Wikipedia Graph}      
	\begin{tabular}{lc}
		\hline
		Specification   & English Wikipedia \\
		\hline
		Amount of vertices ($|V|$)      & 5,251,875 \\
		Amount of Edges ($|E|$)       & 458,867,626 \\
		Average Degree ($\overline{d_{all}}$) & 174 \\
		Median Degree ($\widetilde{d_{all}}$) & 60\\
		Approx. Diameter (D) & 23\\
		Amount of labeled nodes & 4,652,604\\
	\end{tabular}
	\label{tbl:graph}
\end{table}

\subsection{Candidate Generation} Similar to YouTube's deep learning recommender~\cite{Covington:2016:DNN:2959100.2959190}, \textit{WikiRecNet} first generates candidates for a final personalized ranking in a second stage. To generate candidates we first calculate representation vectors (content-based and graph-based) for all articles in our dataset, a process conducted off-line which is presented as \textit{Representation Learning} in Figure \ref{fig:candidate-generation}.  Then, for every user we define her representation as an  aggregation of representation vectors of corresponding articles that were edited by this user. Next, we conduct Nearest Neighbors search with user representation as a \textit{query} in the articles' representation database, a procedure we call \textit{Candidate Generation} and which is conducted online, as shown in Figure \ref{fig:ranking}.

\vspace{12pt}
\noindent
\textbf{Content-based articles representation: Doc2Vec.} For learning the content-based article representation, text features are needed to be extracted first. This can be conducted with traditional document vector space model \cite{salton1975vector} or by using word embeddings such as Word2vec \cite{word2vec} and GloVe \cite{pennington2014glove} and performing an additional step of aggregation. Another option is using directly a full text embedding model and with that goal we use Doc2Vec~\cite{doc2vec}. There are two distinct approaches for learning document embedding with this model. One is Paragraph Vector Distributed Bag-of-Words model (PV-DBOW) model, which is based on word2vec's Continuous Bag-of-Words approach~\cite{word2vec} but instead of word input it accepts paragraph vector and predicts context words for this paragraph. In the second approach, Distributed Memory (PV-DM), which is based on word2vec skip-gram model, the model predicts middle word based on context and paragraph vector given as input. Later on this paper (Section \ref{sec:results}) we show that PV-DBOW is the best fit for our task. We train Doc2Vec-DBOW model on the Corpus of all Wikipedia articles in a given language. Output vectors of Doc2Vec are being passed as input features to the GNC.

\vspace{12pt}
\noindent
\textbf{Graph-based article representation: GraphSAGE.} GraphSAGE has been used as GCN due to its ability to learn with an inductive approach and construct embeddings for unseen nodes. During the pre-processing of the input dataset --snapshot of Wikipedia Dataset-- we create a graph $G(V, E)$ where $V$ denotes the set of articles, and $E$ the set of links between them. GraphSAGE utilizes structural knowledge from graph $G$ and produces new vectors that preserve both text and structural representations.  Due to the inductive nature of GraphSAGE architecture, we do not need to retrain the model every time after adding a new article into the database, this is very important for applying \textit{WikiRecNet} in real scenarios, where new articles are  constantly added~\cite{wikistats2}. 

After producing the document vector and updating the Graph $G$ structure, we can run GraphSAGE model as is, with already trained weights. GCN is a multi-layer network, where each layer can be formulated as:
$$H^{(l+1)} = \sigma(\tilde{D}^{-\frac{1}{2}}\tilde{A}\tilde{D}^{-\frac{1}{2}}H^{(l)}W^{(l+1)}) $$

\noindent
where $\tilde{A} = A + I$ is the  adjacency matrix with self-connections ($I$), $\tilde{D}=\sum_{j} \tilde{A}_{ij}$, $W$ are trainable weights and $H$ is the output of previous layer or $H^{(0)} = X$ is input, $X$  represents node features. An intuitive explanation of this process is that each node collects features of its neighbors that were propagated through trainable filters (convolutions) so called, message passing. On each step node collects knowledge of its neighborhood and propagates its state further on the next step. Thus, properties of 1st, 2nd, ..., nth proximity are being incorporated into node's state along with preserving original features of node's community.

\noindent
\textbf{Optimizing candidate retrieval}. In serving time, recommendation candidates will be produced by applying K-Nearest Neighbors (K-NN) search to find the most similar articles to the user representation vector in the pre-computed database of all articles' representations. K-NN search is one of the main parts of candidate article generation, since its performance in terms of time and resource consumption is very critical for online recommendation in a high-load system. We conducted experiments with different optimizations for K-NN candidate search using FAISS library~\cite{JDH17} : Locality-Sensitive Hashing (LSH), Inverted file with exact post-verification (IVF), Hierarchical Navigable Small World graph exploration (HNSW). Our tests showed that HNSW gives the best speed along with exactly the same recall and MRR as exact search, so with no trade-off in performance we achieved 20x times improvement in speed. Results of these experiments are shown in Table \ref{tab:my_label}.  

\subsection{Ranking of Candidate Articles} 

After learning content and graph-based representations for Wikipedia articles, in the second part of our system  we are trying to model user preferences based on the previous edit history of Wikipedia contributors. With given previous editions and articles, we produce  a relevant a list of candidates, ranked by its relevance for a given user. Our model is  trained on binary labels - \textit{relevant / not relevant} (logistic regression), as shown in Figure \ref{fig:candidate-generation},  but on serving time it will produce probabilities of user interest, which are used as a preference ranking score.

This approach is inspired by Pointwise ranking~\cite{Liu:2009:LRI:1618303.1618304} and is implemented in many similar recommender systems: YouTube\cite{Covington:2016:DNN:2959100.2959190}, eBay\cite{Brovman:2016:OSI:2959100.2959166}. The model is shown on Figure \ref{fig:ranking} and consists of several fully-connected layers with Batch Normalization and ReLU activation after each layer except for the last layer, where sigmoid activation is used. The final model's architecture was selected as follows:
$ 1024 \text{ ReLU -> }  512 \text{ ReLU -> } 256 \text{ ReLU} $. 
As input model accept a concatenated vector of user and candidate representations.

\textbf{Preference score} We define our preference ranking score as the probability that a user $u$ finds a wikipedia article $a_i$ relevant:
\begin{equation}
score(u,a_i) = P(a_i = relevant | u ) = \frac{1}{ 1+ e^{-\Phi(a_i, u) }}    \label{eq:preference score}
\end{equation}
where $u$ represents the user, $a_i$ a candidate wikipedia article, $A$ is the set of all articles to be ranked, and $\Phi(\cdot)$ a weighted sum of the values in the last network layer of the \textit{Candidate ranking} neural network, shown in Figure \ref{fig:ranking}. We train the model with a traditional loss for a binary logistic regression.

%% file: sections/040.tex
\section{Experiments}
\label{sec:Experiments}
We worked with the English version of Wikipedia not only because is the most popular one. In addition, is the most challenging in terms of data processing, and if our system is able to deal with the largest Wikipedia it would be easy to apply later in smaller projects. 

\subsection{Dataset}

All data used  has been downloaded from the official  Wikimedia Dumps~\cite{wikidumps} which  are snapshots of the full Wikipedia.  Some of the objects in the dump (like articles' links) are stored in SQL format, others, with  deeper structure (like articles' text) are stored in XML. 

First of all, for representation learning we built a graph $G(V, E)$, where the set of nodes $V$ is the set of all Wikipedia pages belonging to article namespace\footnote{https://en.wikipedia.org/wiki/Wikipedia:Namespace} and $E$ is the set of directed links between them. The SQL dumps of \emph{page}, \emph{pagelinks}, and \emph{redirects} tables were parsed to organize this data. During pre-processing stage, all links to redirect pages\footnote{https://en.wikipedia.org/wiki/Wikipedia:Redirect} were replaced by their actual destinations. "Category pages", that consists only of links to other pages and do not have their own content, were detected and filtered out. 
We used Apache Spark and GraphX for parallel parsing of SQL dumps and discovering and cleaning Article Graph respectively. The output Graph was converted into binary format with graph-tool~\cite{peixoto_graph-tool_2014} to achieve fast processing (see Table \ref{tbl:graph}). 
For extracting the articles' texts we took the latest revision\footnote{Article's revision is a specific version of article's content after each  \textit{modification}} per each article from XML dump.  We used \textit{Gensim}~\cite{rehurek_lrec} to tokenize and lemmatize text and prepare for the Doc2vec training.

To the facilitate the evaluation in an end-to-end fashion we reorganized the data into a \textit{revisions-per-user} dataset. Only revisions that were created after Jan. 2015 were kept in this dataset, so our recommendations that are based on the latest snapshot of article graph (Oct. 2018) will not recommend too many articles that did not exists on that moment.  

We found that 88\% of contributors are not regular users, since they edited fewer than 5 different articles for selected dates. We also calculated diversity ~\cite{diversity} of users' contribution based on vector representations obtained from Doc2vec. The set of contributors that fits to our needs has mostly edited from 5 to 40 different articles, though diversity of those articles is rather high. That is the main cause our representations cannot be trained against this data like it was done in previous work~\cite{Covington:2016:DNN:2959100.2959190}. Unlike the work by Covington et al.~\cite{Covington:2016:DNN:2959100.2959190}, our training dataset is small (around 60K users) and users' areas of interest are very diverse.

\subsection{Training}
For all training experiments with GraphSAGE we generated a sampled adjacency matrix based on Articles' graph $G$. The width of the adjacency matrix was determined by our experiments, based available memory resources as well as on graph statistics (Table \ref{tbl:graph}). We selected $128$ as maximum amount of neighbors in this matrix. If a node had more than that, then we used random subsampling.
Our GCN architecture consists of two convolutional layers.
On each convolutional step we picked a random sample of $25$ neighbors from this adjacency matrix. This 25-neighbors sample is being resampled on each new batch. For better generalization we used a batch size of 512, since experiments with dropout between convolutional layers led to no improvement in generalization. We set the size for all output vectors to 512 considering a balance between better resolution and available memory.

Document representations from Doc2Vec-DBOW, trained with vector size $300$ and window size $8$, were passed as initial node states and graph edges played the role of labels when the model was trying to predict those edges. We utilized max-margin loss~\cite{Ying2018GraphCN}, as target for training GraphSAGE in link-prediction setup. Model parameters were tuned with the Adam optimizer\cite{DBLP:journals/corr/KingmaB14}.
 \begin{align}
 J(z_u z_i) = E_{v_n \sim P_n(u)} max \{ 0,  z_u \cdot z_{v_u} - z_u \cdot z_i + \Delta \}
 \label{eq:max-margin}
 \end{align}

 For training the ranking model, a dataset from users' history was constructed. As input this model takes 5 articles edited by a user ( representing users' preferences) and 1 candidate that might interest the user. The model tries to predict the probability of relevance of this candidate to the current user. Those 6 input articles are passed through an \textit{Embedding Layer} populated with representations received from GraphSAGE and then concatenated into one vector. We chose positive candidates from actual user history and generated negative candidates with kNN search on constructed articles' representations. Logistic (binary cross-entropy) regression with class-weights (due to high class imbalance) was used as loss function.
 
 \subsection{Evaluation}
 
 To prepare the evaluation dataset, we subsampled windows of size 10 from user's history (from users that were not previously used for training or testing the Deep Ranking model). Our assumption if that the first 5 articles denoted users' area of interest. To compute a single user vector we took element-wise average of representations from the first 5 articles (GraphSAGE representations). We were trying to predict the rest 5. Algorithm can be expressed as follows:
(i) take first 5 articles per user. Calculate average of their embeddings vectors, output this as the user vector representation, (ii) generate candidates by nearest neighbors search of user representation, (iii) sort candidates according to ranking algorithm and select the top $K$. In our evaluation we compare two ranking techniques: sort by cosine similarity, and sort by probability from Deep Ranking model, and (iv) compare Top-K recommendations with the 5 articles in the test set (from second half of the sampled window).

To measure the results we used several metrics: mean average precision (MAP), normalized discounted cumulative gain (nDCG)\cite{Baltrunas:2010:GRR:1864708.1864733} and Recall@k~\cite{cremonesi2010performance}. We calculate these metrics at high \textit{k} values, $k=50$ an $k=100$. Unlike traditional research on top-k recommendation systems usually focusing on small \textit{k} values (\textit{k}=10,20,30), we are specially interested in preventing popularity bias, i.e., having WikiRecNet biased to recommend mostly popular items. Valcarce et al.~\cite{valcarce2018robustness} showed recently that usual top-k ranking metrics measured at higher values of \textit{k} (50, 100) are specially robust to popularity bias, and that is why we use them here.

%% file: sections/050.tex
\setlength\tabcolsep{3pt}.
\begin{table*}[!t]
\centering
\caption{Offline evaluation of generated recommendations on the task of predicting next 5 articles edited by user with percentage improvement over content-based model Doc2Vec (mean-pool) with cosine similarity.}
\begin{tabular}{l l l c c c l l l}
\hline
& & & \multicolumn{3}{c}{K=50} & \multicolumn{3}{c}{K=100}\\ 
\cline{4-6} \cline{7-9}
Model & Aggregate & Rank   & MAP & nDCG & Recall & MAP & nDCG & Recall \\
\hline
WikiRecNet & mean & cosine & 0.0221 & 0.1361 & 0.0846 & 0.0238 (+78\%) & 0.1468 (+66\%) & 0.1179 (+99\%) \\
 & mean & deep-rank & 0.0228 & 0.1363 & 0.0841 & 0.0243 (+82\%) & 0.1493 (+70\%) & 0.1134 (+92\%) \\
 & max & cosine & 0.0192 & 0.1196 & 0.0672 & 0.0206 (+54 \%) & 0.1299 (+47\%) & 0.0923 (+56\%) \\
 & merge & cosine & 0.0208 & 0.1412 & 0.0825 & 0.0227 (+70\%) & 0.1538 (+75\%) & 0.1175 (+99\%) \\
 & merge & deep-rank & \textbf{0.0262} & \textbf{0.1625} & \textbf{0.0935} & \textbf{0.0282 (+111\%)} & \textbf{0.1760 (+100\%)} & \textbf{0.1302 (+120\%)} \\
Doc2Vec & merge & cosine & 0.0085 & 0.0805 & 0.0438 & 0.0092 & 0.0883 & 0.0600 \\
 & mean & cosine & 0.0126 & 0.0821 & 0.0436 & 0.0133 & 0.0880 & 0.0590 \\ 
BM25 & & & 0.0251 & 0.1602 & 0.0921 & 0.0273 & 0.1710 & 0.1290 \\
ALS MF & & & 0.0027 & 0.0163 & 0.044 & 0.0063 & 0.0204 & 0.0609 \\
\hline
\end{tabular}
\label{tbl:results}
\end{table*}

\section{Results}
\label{sec:results}

Results of the evaluation are presented in Table \ref{tbl:results}. We first describe the competing methods:
 

\textbf{Baselines}. We  used two well established methods. The first one is BM25\cite{Robertson:2009:PRF:1704809.1704810}, a probabilistic method used on information retrieval but also applied for content-based filtering in the area of recommendation~\cite{parra2009collaborative}. A second baseline is implicit feedback collaborative filtering optimized with Alternative Least Squares (ALS)~\cite{Hu08collaborativefiltering}. 


\textbf{K-NN recommender}. In addition, we implemented a simple K-NN recommender where the Wikipedia articles are represented by the Doc2vec vector embeddings. Each user $u$ is represented by the articles she has edited, and we test two forms of aggregation to represent the user model: merging the user-edited articles (\textit{merge}) and calculating the mean at each dimension of the document (\textit{mean-pool}). We rank recommended articles by cosine similarity.


\textbf{Aggregations}. Finally,\textit{WikiRecNet} is presented in 5 versions by varying the type of aggregation of articles to represent the user model (merge, mean-pool, max-pool), as well as the method for ranking (cosine similarity and Deep-Rank).

The results in Table \ref{tbl:results} show that WikiRecNet,   using \textit{merge} aggregation and \textit{Deep-Rank} ranking, outperforms the other methods in all metrics. We highlight the following the aspects in the evaluation:

\begin{itemize}
    \item ALS implicit feedback collaborative filtering performs the worst among all methods. This result must be due to the extreme high sparsity of the dataset.
    \item BM25, despite being a simple and traditional content-based filtering method, performs well and remains very competitive.
    \item The simple K-NN based on Doc2Vec representation performs better than ALS, and mean-pool reports better results than merge but only at higher ranking positions (MAP@50, nDCG@50, Recall@50).
    \item Among the WikiRecNet variations, the max-pool aggregation seems to be the least helpful. In terms of nDCG@50 and nDCG@100 (the metric most robust to popularity bias ~\cite{valcarce2018robustness}), merge aggregation seems more effective than mean-pool, and then the combination with DeepRank produce the best performance, with a 100\% increase compared to the Doc2vec mean-pool reference method. 
\end{itemize}

%% file: sections/060.tex
\section{Conclusion}
\label{sec:conclusion}

In this article we have introduced \textit{WikiRecNet}, a neural-based model which aims at recommending Wikipedia articles to editors, in order to help them dealing with the sheer volume of potential articles that might need their attention.  Our approach uses representation learning, i.e., finding alternative ways to represent the Wikipedia articles in order to produce a useful recommendation  without requiring more information than the previous articles edited by targeted users. For this purpose, we used Doc2Vec~\cite{doc2vec} for a content-based representation and GraphSage~\cite{graphsage}, a graph convolutional network, for a graph-based representation.

\textit{WikiRecNet} architecture is composed of two networks, a candidate generation network and a ranking network, and our implementation is able to deal with larges volumes of data, improving existing implementations that were not capable to work in such scenarios. Also, our approach does not need to be retrained when new items are added, facilitating its application in dynamic environments such as Wikipedia. To best of our knowledge, this is the first recommender system especially designed for Wikipedia editors that takes in account such applications constrains, and therefore, can be applied in real world scenarios.

In order to contribute to the community, we provide our code and the graph embedding of each Wikipedia page used in this experiment\footnote{Embeddings in other languages would be also available under request.} available in a public repository, as well as a working demo that can be tested by the Wikipedia editors community\footnote{\url{https://github.com/digitalTranshumant/WikiRecNet-ComplexRec2020}}. With respect to text embeddings, there have been important progresses in the latest years, so another idea for future work will be testing models like BERT \cite{devlin2018bert} or XLNet \cite{yang2019xlnet}.

\section{Acknowledgments}
The author Denis Parra has been funded by the Millennium Institute for Foundational Research on Data (IMFD) and by the Chilean research agency ANID, FONDECYT grant 1191791.

%% file: sample-sigconf.bbl

\begin{thebibliography}{35}


\ifx \showCODEN    \undefined \def \showCODEN     #1{\unskip}     \fi
\ifx \showDOI      \undefined \def \showDOI       #1{#1}\fi
\ifx \showISBNx    \undefined \def \showISBNx     #1{\unskip}     \fi
\ifx \showISBNxiii \undefined \def \showISBNxiii  #1{\unskip}     \fi
\ifx \showISSN     \undefined \def \showISSN      #1{\unskip}     \fi
\ifx \showLCCN     \undefined \def \showLCCN      #1{\unskip}     \fi
\ifx \shownote     \undefined \def \shownote      #1{#1}          \fi
\ifx \showarticletitle \undefined \def \showarticletitle #1{#1}   \fi
\ifx \showURL      \undefined \def \showURL       {\relax}        \fi
\providecommand\bibfield[2]{#2}
\providecommand\bibinfo[2]{#2}
\providecommand\natexlab[1]{#1}
\providecommand\showeprint[2][]{arXiv:#2}

\bibitem[\protect\citeauthoryear{Baltrunas, Makcinskas, and Ricci}{Baltrunas
  et~al\mbox{.}}{2010}]%
        {Baltrunas:2010:GRR:1864708.1864733}
\bibfield{author}{\bibinfo{person}{Linas Baltrunas}, \bibinfo{person}{Tadas
  Makcinskas}, {and} \bibinfo{person}{Francesco Ricci}.}
  \bibinfo{year}{2010}\natexlab{}.
\newblock \showarticletitle{Group Recommendations with Rank Aggregation and
  Collaborative Filtering}. In \bibinfo{booktitle}{\emph{Proceedings of the
  Fourth ACM Conference on Recommender Systems}} \emph{(\bibinfo{series}{RecSys
  '10})}. \bibinfo{publisher}{ACM}, \bibinfo{address}{New York, NY, USA},
  \bibinfo{pages}{119--126}.
\newblock
\showISBNx{978-1-60558-906-0}
\urldef\tempurl%
\url{https://doi.org/10.1145/1864708.1864733}
\showDOI{\tempurl}


\bibitem[\protect\citeauthoryear{Bengio, Courville, and Vincent}{Bengio
  et~al\mbox{.}}{2013}]%
        {bengio2013representation}
\bibfield{author}{\bibinfo{person}{Yoshua Bengio}, \bibinfo{person}{Aaron
  Courville}, {and} \bibinfo{person}{Pascal Vincent}.}
  \bibinfo{year}{2013}\natexlab{}.
\newblock \showarticletitle{Representation learning: A review and new
  perspectives}.
\newblock \bibinfo{journal}{\emph{IEEE transactions on pattern analysis and
  machine intelligence}} \bibinfo{volume}{35}, \bibinfo{number}{8}
  (\bibinfo{year}{2013}), \bibinfo{pages}{1798--1828}.
\newblock


\bibitem[\protect\citeauthoryear{Bobadilla, Serradilla, and Bernal}{Bobadilla
  et~al\mbox{.}}{2010}]%
        {diversity}
\bibfield{author}{\bibinfo{person}{J Bobadilla}, \bibinfo{person}{Francisco
  Serradilla}, {and} \bibinfo{person}{J Bernal}.}
  \bibinfo{year}{2010}\natexlab{}.
\newblock \showarticletitle{A new collaborative filtering metric that improves
  the behavior of recommender systems}.
\newblock \bibinfo{journal}{\emph{Knowledge-Based Systems}}
  \bibinfo{volume}{23} (\bibinfo{date}{08} \bibinfo{year}{2010}),
  \bibinfo{pages}{520--528}.
\newblock
\urldef\tempurl%
\url{https://doi.org/10.1016/j.knosys.2010.03.009}
\showDOI{\tempurl}


\bibitem[\protect\citeauthoryear{Brovman, Jacob, Srinivasan, Neola, Galron,
  Snyder, and Wang}{Brovman et~al\mbox{.}}{2016}]%
        {Brovman:2016:OSI:2959100.2959166}
\bibfield{author}{\bibinfo{person}{Yuri~M. Brovman}, \bibinfo{person}{Marie
  Jacob}, \bibinfo{person}{Natraj Srinivasan}, \bibinfo{person}{Stephen Neola},
  \bibinfo{person}{Daniel Galron}, \bibinfo{person}{Ryan Snyder}, {and}
  \bibinfo{person}{Paul Wang}.} \bibinfo{year}{2016}\natexlab{}.
\newblock \showarticletitle{Optimizing Similar Item Recommendations in a
  Semi-structured Marketplace to Maximize Conversion}. In
  \bibinfo{booktitle}{\emph{Proceedings of the 10th ACM Conference on
  Recommender Systems}} \emph{(\bibinfo{series}{RecSys '16})}.
  \bibinfo{publisher}{ACM}, \bibinfo{address}{New York, NY, USA},
  \bibinfo{pages}{199--202}.
\newblock
\showISBNx{978-1-4503-4035-9}
\urldef\tempurl%
\url{https://doi.org/10.1145/2959100.2959166}
\showDOI{\tempurl}


\bibitem[\protect\citeauthoryear{Choi, Alexander, Kraut, and Levine}{Choi
  et~al\mbox{.}}{2010}]%
        {newcomers1}
\bibfield{author}{\bibinfo{person}{Boreum Choi}, \bibinfo{person}{Kira
  Alexander}, \bibinfo{person}{Robert~E Kraut}, {and} \bibinfo{person}{John~M
  Levine}.} \bibinfo{year}{2010}\natexlab{}.
\newblock \showarticletitle{Socialization tactics in wikipedia and their
  effects}. In \bibinfo{booktitle}{\emph{Proceedings of the 2010 ACM conference
  on Computer supported cooperative work}}. ACM, \bibinfo{pages}{107--116}.
\newblock


\bibitem[\protect\citeauthoryear{Consonni, Laniado, and Montresor}{Consonni
  et~al\mbox{.}}{2019}]%
        {consonni2019wikilinkgraphs}
\bibfield{author}{\bibinfo{person}{Cristian Consonni}, \bibinfo{person}{David
  Laniado}, {and} \bibinfo{person}{Alberto Montresor}.}
  \bibinfo{year}{2019}\natexlab{}.
\newblock \showarticletitle{WikiLinkGraphs: A complete, longitudinal and
  multi-language dataset of the Wikipedia link networks}.
\newblock \bibinfo{journal}{\emph{arXiv preprint arXiv:1902.04298}}
  (\bibinfo{year}{2019}).
\newblock


\bibitem[\protect\citeauthoryear{Covington, Adams, and Sargin}{Covington
  et~al\mbox{.}}{2016}]%
        {Covington:2016:DNN:2959100.2959190}
\bibfield{author}{\bibinfo{person}{Paul Covington}, \bibinfo{person}{Jay
  Adams}, {and} \bibinfo{person}{Emre Sargin}.}
  \bibinfo{year}{2016}\natexlab{}.
\newblock \showarticletitle{Deep Neural Networks for YouTube Recommendations}.
  In \bibinfo{booktitle}{\emph{Proceedings of the 10th ACM Conference on
  Recommender Systems}} \emph{(\bibinfo{series}{RecSys '16})}.
  \bibinfo{publisher}{ACM}, \bibinfo{address}{New York, NY, USA},
  \bibinfo{pages}{191--198}.
\newblock
\showISBNx{978-1-4503-4035-9}
\urldef\tempurl%
\url{https://doi.org/10.1145/2959100.2959190}
\showDOI{\tempurl}


\bibitem[\protect\citeauthoryear{Cremonesi, Koren, and Turrin}{Cremonesi
  et~al\mbox{.}}{2010}]%
        {cremonesi2010performance}
\bibfield{author}{\bibinfo{person}{Paolo Cremonesi}, \bibinfo{person}{Yehuda
  Koren}, {and} \bibinfo{person}{Roberto Turrin}.}
  \bibinfo{year}{2010}\natexlab{}.
\newblock \showarticletitle{Performance of recommender algorithms on top-n
  recommendation tasks}. In \bibinfo{booktitle}{\emph{Proceedings of the fourth
  ACM conference on Recommender systems}}. ACM, \bibinfo{pages}{39--46}.
\newblock


\bibitem[\protect\citeauthoryear{Devlin, Chang, Lee, and Toutanova}{Devlin
  et~al\mbox{.}}{2018}]%
        {devlin2018bert}
\bibfield{author}{\bibinfo{person}{Jacob Devlin}, \bibinfo{person}{Ming-Wei
  Chang}, \bibinfo{person}{Kenton Lee}, {and} \bibinfo{person}{Kristina
  Toutanova}.} \bibinfo{year}{2018}\natexlab{}.
\newblock \showarticletitle{Bert: Pre-training of deep bidirectional
  transformers for language understanding}.
\newblock \bibinfo{journal}{\emph{arXiv preprint arXiv:1810.04805}}
  (\bibinfo{year}{2018}).
\newblock


\bibitem[\protect\citeauthoryear{Duvenaud, Maclaurin, Iparraguirre, Bombarell,
  Hirzel, Aspuru-Guzik, and Adams}{Duvenaud et~al\mbox{.}}{2015}]%
        {NIPS2015_5954}
\bibfield{author}{\bibinfo{person}{David~K Duvenaud}, \bibinfo{person}{Dougal
  Maclaurin}, \bibinfo{person}{Jorge Iparraguirre}, \bibinfo{person}{Rafael
  Bombarell}, \bibinfo{person}{Timothy Hirzel}, \bibinfo{person}{Alan
  Aspuru-Guzik}, {and} \bibinfo{person}{Ryan~P Adams}.}
  \bibinfo{year}{2015}\natexlab{}.
\newblock \showarticletitle{Convolutional Networks on Graphs for Learning
  Molecular Fingerprints}.
\newblock In \bibinfo{booktitle}{\emph{Advances in Neural Information
  Processing Systems 28}}, \bibfield{editor}{\bibinfo{person}{C.~Cortes},
  \bibinfo{person}{N.~D. Lawrence}, \bibinfo{person}{D.~D. Lee},
  \bibinfo{person}{M.~Sugiyama}, {and} \bibinfo{person}{R.~Garnett}} (Eds.).
  \bibinfo{publisher}{Curran Associates, Inc.}, \bibinfo{pages}{2224--2232}.
\newblock
\urldef\tempurl%
\url{http://papers.nips.cc/paper/5954-convolutional-networks-on-graphs-for-learning-molecular-fingerprints.pdf}
\showURL{%
\tempurl}


\bibitem[\protect\citeauthoryear{Foundation}{Foundation}{2018}]%
        {wikidumps}
\bibfield{author}{\bibinfo{person}{Wikimedia Foundation}.}
  \bibinfo{year}{2018}\natexlab{}.
\newblock \bibinfo{title}{Wikimedia Downloads}.
\newblock
\newblock
\urldef\tempurl%
\url{https://dumps.wikimedia.org}
\showURL{%
\tempurl}
\newblock
\shownote{[Online; accessed 14. Oct. 2019].}


\bibitem[\protect\citeauthoryear{Foundation}{Foundation}{2019}]%
        {wikistats2}
\bibfield{author}{\bibinfo{person}{Wikimedia Foundation}.}
  \bibinfo{year}{2019}\natexlab{}.
\newblock \bibinfo{title}{{Wikimedia Statistics - All wikis}}.
\newblock
\newblock
\urldef\tempurl%
\url{https://stats.wikimedia.org/v2/#/all-projects}
\showURL{%
\tempurl}
\newblock
\shownote{[Online; accessed 13. Oct. 2019].}


\bibitem[\protect\citeauthoryear{Grover and Leskovec}{Grover and
  Leskovec}{2016}]%
        {GroverL16}
\bibfield{author}{\bibinfo{person}{Aditya Grover} {and} \bibinfo{person}{Jure
  Leskovec}.} \bibinfo{year}{2016}\natexlab{}.
\newblock \showarticletitle{node2vec: Scalable Feature Learning for Networks.}
\newblock \bibinfo{journal}{\emph{CoRR}}  \bibinfo{volume}{abs/1607.00653}
  (\bibinfo{year}{2016}).
\newblock
\urldef\tempurl%
\url{http://dblp.uni-trier.de/db/journals/corr/corr1607.html#GroverL16}
\showURL{%
\tempurl}


\bibitem[\protect\citeauthoryear{Hamilton, Ying, and Leskovec}{Hamilton
  et~al\mbox{.}}{2017}]%
        {graphsage}
\bibfield{author}{\bibinfo{person}{William~L. Hamilton}, \bibinfo{person}{Rex
  Ying}, {and} \bibinfo{person}{Jure Leskovec}.}
  \bibinfo{year}{2017}\natexlab{}.
\newblock \showarticletitle{Inductive Representation Learning on Large Graphs}.
  In \bibinfo{booktitle}{\emph{NIPS}}.
\newblock


\bibitem[\protect\citeauthoryear{Hu, Koren, and Volinsky}{Hu
  et~al\mbox{.}}{2008}]%
        {Hu08collaborativefiltering}
\bibfield{author}{\bibinfo{person}{Yifan Hu}, \bibinfo{person}{Yehuda Koren},
  {and} \bibinfo{person}{Chris Volinsky}.} \bibinfo{year}{2008}\natexlab{}.
\newblock \showarticletitle{Collaborative filtering for implicit feedback
  datasets}. In \bibinfo{booktitle}{\emph{In IEEE International Conference on
  Data Mining (ICDM 2008}}. \bibinfo{pages}{263--272}.
\newblock


\bibitem[\protect\citeauthoryear{Johnson, Douze, and J{\'e}gou}{Johnson
  et~al\mbox{.}}{2017}]%
        {JDH17}
\bibfield{author}{\bibinfo{person}{Jeff Johnson}, \bibinfo{person}{Matthijs
  Douze}, {and} \bibinfo{person}{Herv{\'e} J{\'e}gou}.}
  \bibinfo{year}{2017}\natexlab{}.
\newblock \showarticletitle{Billion-scale similarity search with GPUs}.
\newblock \bibinfo{journal}{\emph{arXiv preprint arXiv:1702.08734}}
  (\bibinfo{year}{2017}).
\newblock


\bibitem[\protect\citeauthoryear{Kingma and Ba}{Kingma and Ba}{2015}]%
        {DBLP:journals/corr/KingmaB14}
\bibfield{author}{\bibinfo{person}{Diederik~P. Kingma} {and}
  \bibinfo{person}{Jimmy Ba}.} \bibinfo{year}{2015}\natexlab{}.
\newblock \showarticletitle{Adam: {A} Method for Stochastic Optimization}. In
  \bibinfo{booktitle}{\emph{3rd International Conference on Learning
  Representations, {ICLR} 2015, San Diego, CA, USA, May 7-9, 2015, Conference
  Track Proceedings}}.
\newblock
\urldef\tempurl%
\url{http://arxiv.org/abs/1412.6980}
\showURL{%
\tempurl}


\bibitem[\protect\citeauthoryear{Kipf and Welling}{Kipf and Welling}{2016}]%
        {gcn}
\bibfield{author}{\bibinfo{person}{Thomas~N. Kipf} {and} \bibinfo{person}{Max
  Welling}.} \bibinfo{year}{2016}\natexlab{}.
\newblock \showarticletitle{Semi-Supervised Classification with Graph
  Convolutional Networks}.
\newblock \bibinfo{journal}{\emph{CoRR}}  \bibinfo{volume}{abs/1609.02907}
  (\bibinfo{year}{2016}).
\newblock


\bibitem[\protect\citeauthoryear{Le and Mikolov}{Le and Mikolov}{2014}]%
        {doc2vec}
\bibfield{author}{\bibinfo{person}{Quoc~V. Le} {and} \bibinfo{person}{Tomas
  Mikolov}.} \bibinfo{year}{2014}\natexlab{}.
\newblock \showarticletitle{Distributed Representations of Sentences and
  Documents}.
\newblock \bibinfo{journal}{\emph{CoRR}}  \bibinfo{volume}{abs/1405.4053}
  (\bibinfo{year}{2014}).
\newblock
\showeprint[arxiv]{1405.4053}
\urldef\tempurl%
\url{http://arxiv.org/abs/1405.4053}
\showURL{%
\tempurl}


\bibitem[\protect\citeauthoryear{Liu}{Liu}{2009}]%
        {Liu:2009:LRI:1618303.1618304}
\bibfield{author}{\bibinfo{person}{Tie-Yan Liu}.}
  \bibinfo{year}{2009}\natexlab{}.
\newblock \showarticletitle{Learning to Rank for Information Retrieval}.
\newblock \bibinfo{journal}{\emph{Found. Trends Inf. Retr.}}
  \bibinfo{volume}{3}, \bibinfo{number}{3} (\bibinfo{date}{March}
  \bibinfo{year}{2009}), \bibinfo{pages}{225--331}.
\newblock
\showISSN{1554-0669}
\urldef\tempurl%
\url{https://doi.org/10.1561/1500000016}
\showDOI{\tempurl}


\bibitem[\protect\citeauthoryear{Mikolov, Chen, Corrado, and Dean}{Mikolov
  et~al\mbox{.}}{2013}]%
        {word2vec}
\bibfield{author}{\bibinfo{person}{Tomas Mikolov}, \bibinfo{person}{Kai Chen},
  \bibinfo{person}{Greg Corrado}, {and} \bibinfo{person}{Jeffrey Dean}.}
  \bibinfo{year}{2013}\natexlab{}.
\newblock \showarticletitle{Efficient Estimation of Word Representations in
  Vector Space}.
\newblock \bibinfo{journal}{\emph{CoRR}}  \bibinfo{volume}{abs/1301.3781}
  (\bibinfo{year}{2013}).
\newblock
\showeprint[arxiv]{1301.3781}
\urldef\tempurl%
\url{http://arxiv.org/abs/1301.3781}
\showURL{%
\tempurl}


\bibitem[\protect\citeauthoryear{Morgan, Bouterse, Walls, and Stierch}{Morgan
  et~al\mbox{.}}{2013}]%
        {newcomers2}
\bibfield{author}{\bibinfo{person}{Jonathan~T Morgan}, \bibinfo{person}{Siko
  Bouterse}, \bibinfo{person}{Heather Walls}, {and} \bibinfo{person}{Sarah
  Stierch}.} \bibinfo{year}{2013}\natexlab{}.
\newblock \showarticletitle{Tea and sympathy: crafting positive new user
  experiences on wikipedia}. In \bibinfo{booktitle}{\emph{Proceedings of the
  2013 conference on Computer supported cooperative work}}. ACM,
  \bibinfo{pages}{839--848}.
\newblock


\bibitem[\protect\citeauthoryear{Parra and Brusilovsky}{Parra and
  Brusilovsky}{2009}]%
        {parra2009collaborative}
\bibfield{author}{\bibinfo{person}{Denis Parra} {and} \bibinfo{person}{Peter
  Brusilovsky}.} \bibinfo{year}{2009}\natexlab{}.
\newblock \showarticletitle{Collaborative filtering for social tagging systems:
  an experiment with CiteULike}. In \bibinfo{booktitle}{\emph{Proceedings of
  the third ACM conference on Recommender systems}}. ACM,
  \bibinfo{pages}{237--240}.
\newblock


\bibitem[\protect\citeauthoryear{Peixoto}{Peixoto}{2014}]%
        {peixoto_graph-tool_2014}
\bibfield{author}{\bibinfo{person}{Tiago~P. Peixoto}.}
  \bibinfo{year}{2014}\natexlab{}.
\newblock \showarticletitle{The graph-tool python library}.
\newblock \bibinfo{journal}{\emph{figshare}} (\bibinfo{year}{2014}).
\newblock
\urldef\tempurl%
\url{https://doi.org/10.6084/m9.figshare.1164194}
\showDOI{\tempurl}


\bibitem[\protect\citeauthoryear{Pennington, Socher, and Manning}{Pennington
  et~al\mbox{.}}{2014}]%
        {pennington2014glove}
\bibfield{author}{\bibinfo{person}{Jeffrey Pennington},
  \bibinfo{person}{Richard Socher}, {and} \bibinfo{person}{Christopher
  Manning}.} \bibinfo{year}{2014}\natexlab{}.
\newblock \showarticletitle{Glove: Global vectors for word representation}. In
  \bibinfo{booktitle}{\emph{Proceedings of the 2014 conference on empirical
  methods in natural language processing (EMNLP)}}.
  \bibinfo{pages}{1532--1543}.
\newblock


\bibitem[\protect\citeauthoryear{Piccardi, Catasta, Zia, and West}{Piccardi
  et~al\mbox{.}}{2018}]%
        {piccardi2018structuring}
\bibfield{author}{\bibinfo{person}{Tiziano Piccardi}, \bibinfo{person}{Michele
  Catasta}, \bibinfo{person}{Leila Zia}, {and} \bibinfo{person}{Robert West}.}
  \bibinfo{year}{2018}\natexlab{}.
\newblock \showarticletitle{Structuring Wikipedia Articles with Section
  Recommendations}.
\newblock \bibinfo{journal}{\emph{arXiv preprint arXiv:1804.05995}}
  (\bibinfo{year}{2018}).
\newblock


\bibitem[\protect\citeauthoryear{{\v R}eh{\r u}{\v r}ek and Sojka}{{\v R}eh{\r
  u}{\v r}ek and Sojka}{2010}]%
        {rehurek_lrec}
\bibfield{author}{\bibinfo{person}{Radim {\v R}eh{\r u}{\v r}ek} {and}
  \bibinfo{person}{Petr Sojka}.} \bibinfo{year}{2010}\natexlab{}.
\newblock \showarticletitle{{Software Framework for Topic Modelling with Large
  Corpora}}. In \bibinfo{booktitle}{\emph{{Proceedings of the LREC 2010
  Workshop on New Challenges for NLP Frameworks}}}. \bibinfo{publisher}{ELRA},
  \bibinfo{address}{Valletta, Malta}, \bibinfo{pages}{45--50}.
\newblock
\newblock
\shownote{\url{http://is.muni.cz/publication/884893/en}.}


\bibitem[\protect\citeauthoryear{Robertson and Zaragoza}{Robertson and
  Zaragoza}{2009}]%
        {Robertson:2009:PRF:1704809.1704810}
\bibfield{author}{\bibinfo{person}{Stephen Robertson} {and}
  \bibinfo{person}{Hugo Zaragoza}.} \bibinfo{year}{2009}\natexlab{}.
\newblock \showarticletitle{The Probabilistic Relevance Framework: BM25 and
  Beyond}.
\newblock \bibinfo{journal}{\emph{Found. Trends Inf. Retr.}}
  \bibinfo{volume}{3}, \bibinfo{number}{4} (\bibinfo{date}{April}
  \bibinfo{year}{2009}), \bibinfo{pages}{333--389}.
\newblock
\showISSN{1554-0669}
\urldef\tempurl%
\url{https://doi.org/10.1561/1500000019}
\showDOI{\tempurl}


\bibitem[\protect\citeauthoryear{Salton, Wong, and Yang}{Salton
  et~al\mbox{.}}{1975}]%
        {salton1975vector}
\bibfield{author}{\bibinfo{person}{Gerard Salton}, \bibinfo{person}{Anita
  Wong}, {and} \bibinfo{person}{Chung-Shu Yang}.}
  \bibinfo{year}{1975}\natexlab{}.
\newblock \showarticletitle{A vector space model for automatic indexing}.
\newblock \bibinfo{journal}{\emph{Commun. ACM}} \bibinfo{volume}{18},
  \bibinfo{number}{11} (\bibinfo{year}{1975}), \bibinfo{pages}{613--620}.
\newblock


\bibitem[\protect\citeauthoryear{Schein, Popescul, Ungar, and Pennock}{Schein
  et~al\mbox{.}}{2002}]%
        {schein2002methods}
\bibfield{author}{\bibinfo{person}{Andrew~I Schein},
  \bibinfo{person}{Alexandrin Popescul}, \bibinfo{person}{Lyle~H Ungar}, {and}
  \bibinfo{person}{David~M Pennock}.} \bibinfo{year}{2002}\natexlab{}.
\newblock \showarticletitle{Methods and metrics for cold-start
  recommendations}. In \bibinfo{booktitle}{\emph{Proceedings of the 25th annual
  international ACM SIGIR conference on Research and development in information
  retrieval}}. ACM, \bibinfo{pages}{253--260}.
\newblock


\bibitem[\protect\citeauthoryear{Valcarce, Bellog{\'\i}n, Parapar, and
  Castells}{Valcarce et~al\mbox{.}}{2018}]%
        {valcarce2018robustness}
\bibfield{author}{\bibinfo{person}{Daniel Valcarce}, \bibinfo{person}{Alejandro
  Bellog{\'\i}n}, \bibinfo{person}{Javier Parapar}, {and}
  \bibinfo{person}{Pablo Castells}.} \bibinfo{year}{2018}\natexlab{}.
\newblock \showarticletitle{On the robustness and discriminative power of
  information retrieval metrics for top-N recommendation}. In
  \bibinfo{booktitle}{\emph{Proceedings of the 12th ACM Conference on
  Recommender Systems}}. ACM, \bibinfo{pages}{260--268}.
\newblock


\bibitem[\protect\citeauthoryear{Wulczyn, West, Zia, and Leskovec}{Wulczyn
  et~al\mbox{.}}{2016}]%
        {wulczyn2016growing}
\bibfield{author}{\bibinfo{person}{Ellery Wulczyn}, \bibinfo{person}{Robert
  West}, \bibinfo{person}{Leila Zia}, {and} \bibinfo{person}{Jure Leskovec}.}
  \bibinfo{year}{2016}\natexlab{}.
\newblock \showarticletitle{Growing wikipedia across languages via
  recommendation}. In \bibinfo{booktitle}{\emph{Proceedings of the 25th
  International Conference on World Wide Web}}. International World Wide Web
  Conferences Steering Committee, \bibinfo{pages}{975--985}.
\newblock


\bibitem[\protect\citeauthoryear{Yang, Dai, Yang, Carbonell, Salakhutdinov, and
  Le}{Yang et~al\mbox{.}}{2019}]%
        {yang2019xlnet}
\bibfield{author}{\bibinfo{person}{Zhilin Yang}, \bibinfo{person}{Zihang Dai},
  \bibinfo{person}{Yiming Yang}, \bibinfo{person}{Jaime Carbonell},
  \bibinfo{person}{Ruslan Salakhutdinov}, {and} \bibinfo{person}{Quoc~V Le}.}
  \bibinfo{year}{2019}\natexlab{}.
\newblock \showarticletitle{XLNet: Generalized Autoregressive Pretraining for
  Language Understanding}.
\newblock \bibinfo{journal}{\emph{arXiv preprint arXiv:1906.08237}}
  (\bibinfo{year}{2019}).
\newblock


\bibitem[\protect\citeauthoryear{Ying, He, Chen, Eksombatchai, Hamilton, and
  Leskovec}{Ying et~al\mbox{.}}{2018}]%
        {Ying2018GraphCN}
\bibfield{author}{\bibinfo{person}{Rex Ying}, \bibinfo{person}{Ruining He},
  \bibinfo{person}{Kaifeng Chen}, \bibinfo{person}{Pong Eksombatchai},
  \bibinfo{person}{William~L. Hamilton}, {and} \bibinfo{person}{Jure
  Leskovec}.} \bibinfo{year}{2018}\natexlab{}.
\newblock \showarticletitle{Graph Convolutional Neural Networks for Web-Scale
  Recommender Systems}. In \bibinfo{booktitle}{\emph{KDD}}.
\newblock


\bibitem[\protect\citeauthoryear{Zhou, Niu, and Chen}{Zhou
  et~al\mbox{.}}{2018}]%
        {fastnode2vec}
\bibfield{author}{\bibinfo{person}{Dongyan Zhou}, \bibinfo{person}{Songjie
  Niu}, {and} \bibinfo{person}{Shimin Chen}.} \bibinfo{year}{2018}\natexlab{}.
\newblock \showarticletitle{Efficient Graph Computation for Node2Vec.}
\newblock \bibinfo{journal}{\emph{CoRR}}  \bibinfo{volume}{abs/1805.00280}
  (\bibinfo{year}{2018}).
\newblock
\urldef\tempurl%
\url{http://dblp.uni-trier.de/db/journals/corr/corr1805.html#abs-1805-00280}
\showURL{%
\tempurl}


\end{thebibliography}
